\documentclass[%
 reprint,
superscriptaddress,
 amsmath,amssymb,
 aps,
]{revtex4-2}

\usepackage{orcidlink}
\usepackage{xcolor}
\usepackage{float}
\usepackage{graphicx}
\usepackage{dcolumn}
\usepackage{bm}
\usepackage{hyperref}
\hypersetup{
  colorlinks=true,
  allcolors=blue!60!black, 
  pdfborder={0 0 0},
}
\usepackage{aas_macros}

\begin{document}

\title{Deep-Learning Denoising of Radio Signals for Ultra-High-Energy\\ Cosmic-Ray Detection}

\author{Zhisen Lai\,\orcidlink{0000-0001-8483-9089}} 
\email{zlai@sfsu.edu}
\affiliation{Department of Physics and Astronomy, San Francisco State University, San Francisco, CA 94132, USA}

\author{Oscar Macias\,\orcidlink{0000-0001-8867-2693}}
\email{macias@sfsu.edu}
\affiliation{Department of Physics and Astronomy, San Francisco State University, San Francisco, CA 94132, USA}
\affiliation{GRAPPA -- Gravitational and Astroparticle Physics Amsterdam, University of Amsterdam, Science Park 904, 1098 XH Amsterdam, The Netherlands}

\author{Aur\'elien Benoit-L\'evy}%
\affiliation{Université Paris-Saclay, CEA, List, F-91120, Palaiseau, France}%

\author{Ars\`ene Ferri\`ere}
\affiliation{Université Paris-Saclay, CEA, List, F-91120, Palaiseau, France}
\affiliation{Sorbonne Université, CNRS, Laboratoire de Physique Nucléaire et des Hautes Energies
(LPNHE), 4 Pl. Jussieu, Paris, 75005, France}

\author{Mat\'ias Tueros \orcidlink{0000-0003-1570-1419}}
\affiliation{Instituto de F\'isica La Plata - CCT La Plata - CONICET-UNLP, Diag 113 y 63, La Plata (1900), Argentina} \affiliation{Depto. de Física, Fac. de Cs. Ex., Universidad Nacional de La Plata,  Casilla de Correo 67, La Plata (1900), Argentina}

\date{\today}

\begin{abstract}
Extensive air showers initiated by ultra-high-energy cosmic rays (UHECRs) produce broadband radio pulses detectable over large areas, but Galactic and instrumental backgrounds limit sensitivity near threshold. The Giant Radio Array for Neutrino Detection (GRAND) will use large arrays of autonomous radio antennas to detect these air showers, requiring reliable recovery of weak pulses from background fluctuations. We develop a convolutional denoiser for GRAND-like antenna traces that combines the time-domain waveform with the magnitude and phase of its Fourier transform. The model is trained on $4.1\times10^5$ simulated traces constructed from \textsc{ZHAireS} air-shower emission, a reduced GRAND-like RF-chain response, and broadband noise normalized to the expected Galactic-noise scale. In these simulations, denoising increases the detection probability at fixed false-positive rate and improves pulse-time recovery near threshold. The reconstructed peak amplitude is attenuated in the noise-dominated regime. This bias decreases as the pulse becomes better resolved, but a smaller channel-dependent bias persists at higher SNR. Near threshold, denoising therefore primarily improves candidate recovery and timing, while the measured amplitude response provides a basis for calibrating amplitude-sensitive observables.
\end{abstract}

\maketitle

\section{Introduction}

The flux of ultra-high-energy cosmic rays (UHECRs) is low at the highest
energies, requiring large exposure for their detection~\cite{Anchordoqui:2018qom,
AlvesBatista:2024czs}. The Giant Radio Array for Neutrino
Detection (GRAND) is a planned radio observatory for ultra-high-energy
neutrinos, cosmic rays, and gamma rays~\cite{GRAND:2018iaj}. GRAND will
detect these particles through the coherent radio emission produced by
the extensive air showers they generate, with sensitivity concentrated
in the $50$--$200~\mathrm{MHz}$ band. Near threshold, these short radio
pulses must be identified against ambient and instrumental backgrounds.

Galactic background radiation and human-made radio-frequency interference (RFI) dominate the noise floor at most sites and can mask faint cosmic-ray pulses~\cite{Huege:2016veh, Schroder:2016hrv}. These backgrounds raise the effective detection threshold and reduce sensitivity to weak air showers. Radio detection near threshold therefore requires recovery of short cosmic-ray pulses from contaminated waveforms, with improved effective signal-to-noise ratio (SNR) and a lower detection threshold.

Classical filtering methods have been investigated for this purpose. In particular, matched filtering~\cite{Shipilov:2019} using an expected pulse template can improve sensitivity under idealized noise conditions, such as white noise, and has been shown in simulations to reduce trigger thresholds. However, its performance degrades under realistic non-Gaussian noise because RFI does not satisfy the assumptions of an idealized random background.

Wavelet methods isolate transient structure on several scales and have been applied to narrow-band interference and Gaussian noise~\cite{Watanabe:2021}. Early digital radio detectors also used iterative subtraction based on the radio-astronomy \texttt{CLEAN} algorithm~\cite{Hogbom:1974} together with beamforming~\cite{LOPES:2004akj}. The performance of these approaches depends on the assumed signal and background properties, particularly for complex or nonstationary interference. Analyses therefore often impose high post-filtering SNR thresholds that reject weak air showers~\cite{martinelli:2025}.

\begin{figure*}[t!]
    \centering
    \includegraphics[width=1.0\linewidth]{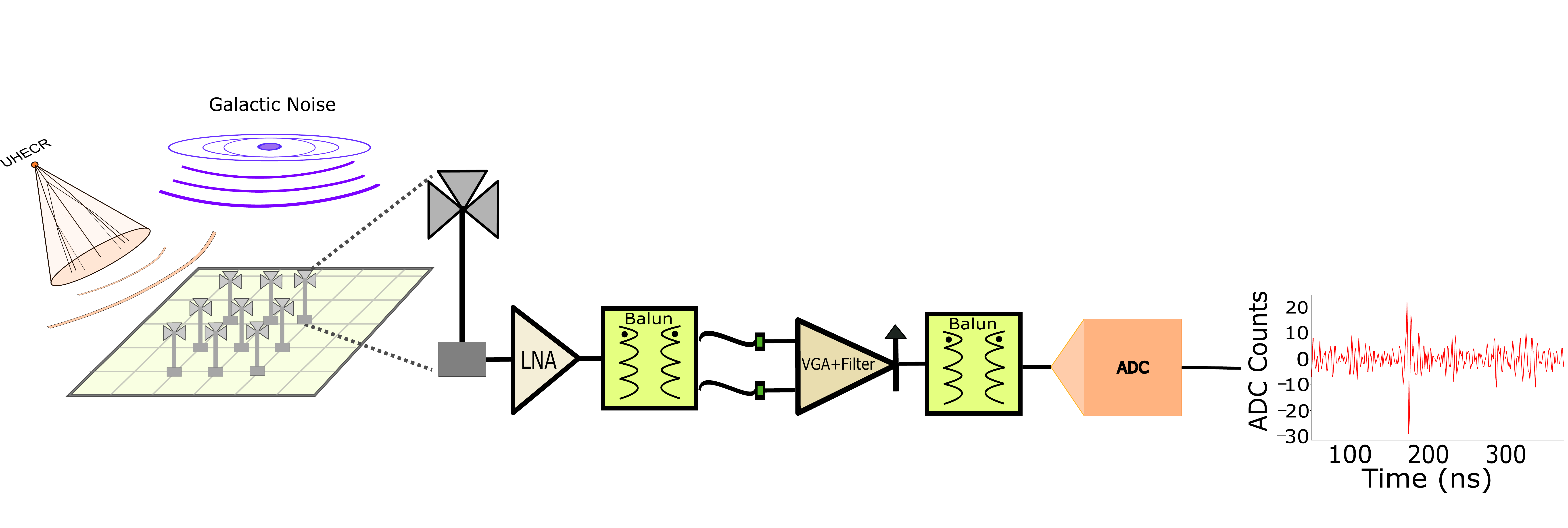}
    \caption{\textbf{Construction of the simulated traces.}
A UHECR air shower produces a short radio pulse at a GRAND
\texttt{HORIZON}-like antenna~\cite{GRAND:2018iaj}.
The simulated pulse is passed through a simplified GRAND-like RF-chain
model that imposes the band limitation and net gain used in this
antenna-level study. This gives the noise-free reference trace.
For the network input, broadband stochastic noise is added before
applying the same response. These processed traces define the
detector-level waveforms used throughout the analysis.}
    \label{fig:pipeline}
\end{figure*}

\begin{figure}[t!]
    \centering
    \includegraphics[width=1.0\linewidth]{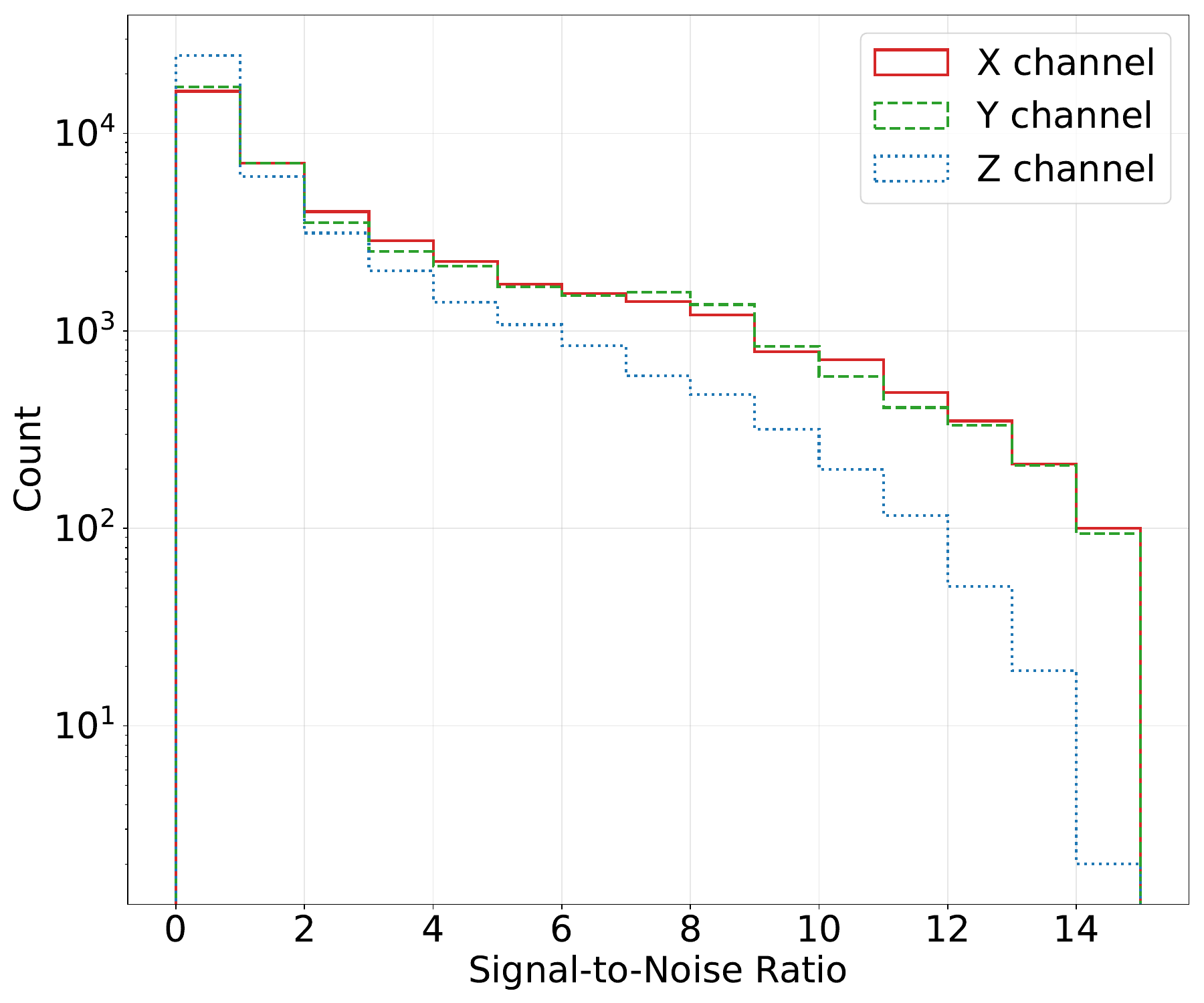}
    \caption{\textbf{Input-SNR distribution of the evaluation sample.} Histogram of $\mathrm{SNR}_{\rm in}$ for the simulated GRAND-like antenna traces in the three polarization channels (X, Y, Z). The input SNR is defined in Eq.~\ref{eq:snr-methods}.}
    \label{fig:training-distribution}
\end{figure}

\begin{figure*}[t!]
    \centering
    \includegraphics[width=1.0\linewidth]{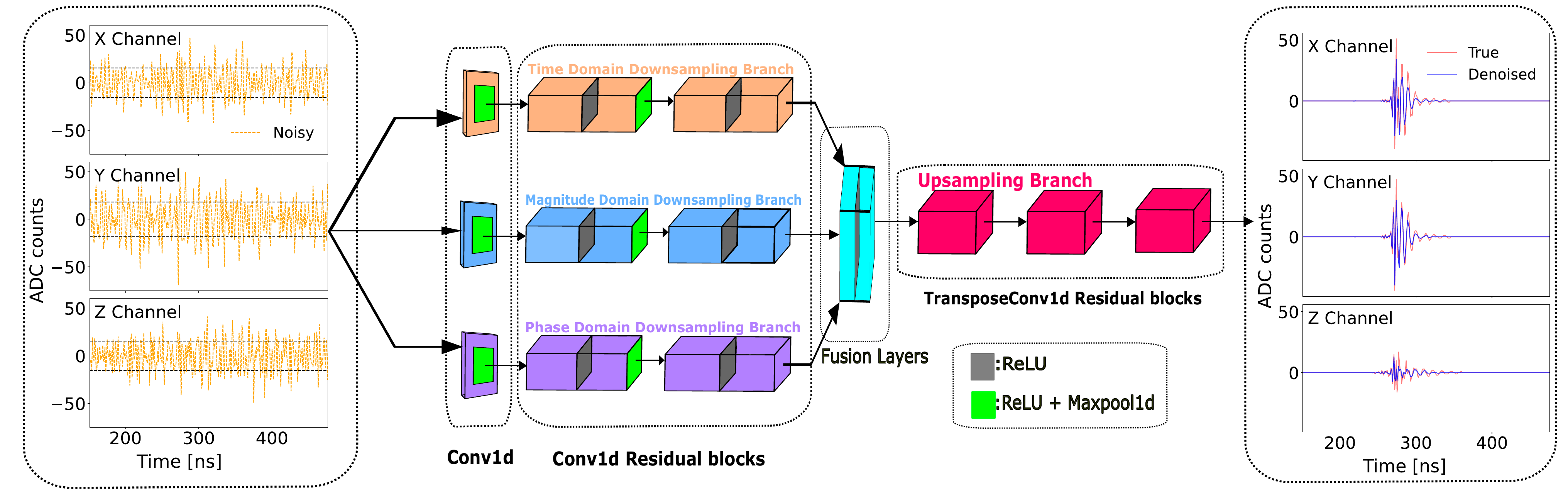}
    \caption{\textbf{Encoder--decoder architecture.} The input contains three polarization channels with 512 samples per channel, extracted from simulated traces stored with 1,024 samples. The encoder contains a time-domain branch (orange) and a Fourier-domain branch that separates magnitude (dark blue) and phase (purple). Each branch begins with a one-dimensional convolution followed by two residual blocks. The fusion module (cyan) combines the three feature maps using two additional convolutional layers. The decoder (red) returns denoised traces with the same dimensions as the input.}
    \label{fig:Model}
\end{figure*}

Learned denoising methods have been applied to signal-reconstruction problems in radio astronomy and gravitational-wave data analysis~\cite{Gheller:2021, Terris:2022, Einig:2023, Reissel:2025ykl}. For air-shower radio data, deep-learning methods have been used to identify weak pulses and reconstruct noise-free waveforms~\cite{Erdmann:2019nie, Bezyazeekov:2021, Schroeder:2024bzf, IceCube:2025swy}. In simulated data, Ref.~\cite{Erdmann:2019nie} reported a pulse-energy resolution of order 20\% without bias and suppression of narrow-band RFI. The Tunka-Rex Collaboration found that an autoencoder lowered the detection threshold in measured data while preserving shower reconstruction~\cite{Bezyazeekov:2021}. An IceCube analysis of several months of data recovered about five times more events than a conventional high-threshold trigger, with improved purity and unchanged angular resolution~\cite{IceCube:2025swy}. For GRAND-like traces, however, timing, waveform fidelity, and amplitude response must be validated separately.

We extend the initial study of Ref.~\cite{Benoit-Levy:2025dpq} using a denoising autoencoder that processes the three polarization traces jointly. The encoder combines local time-domain features with the magnitude and phase of the Fourier transform. During training, each contaminated detector-level trace is paired with its noise-free reference.

We evaluate the model using antenna-level simulations that include a
reduced GRAND-like RF-chain response and broadband stochastic noise
normalized to the expected Galactic-noise scale. These simulations
omit the full local sidereal time (LST) dependence and spectral
structure of the sky background, nonstationary anthropogenic RFI, and
variations in the detector response. Dedicated GRAND studies,
including NUTRIG~\cite{Correa:2025ovu,Chiche:2024ohe}, address
self-triggering and background rejection under more realistic
operating conditions. We restrict the present study to antenna-level
denoising under these controlled conditions. Incorporating the full
detector response and more realistic backgrounds is left to future
studies.

We demonstrate that our proposed denoiser increases the detection probability at fixed false-positive rate and improves pulse-time recovery near threshold for this detector and noise model. At low SNR, however, the network attenuates part of the weak pulse, and a smaller channel-dependent bias remains at higher SNR. We
therefore characterize timing, amplitude response, and waveform fidelity
separately, since denoising affects these observables differently near
threshold.

The remainder of this paper is organized as follows.
Section~\ref{sec:descript-of-simulated-data} describes the simulated
data and preprocessing, and Section~\ref{sec:methods} presents the
denoiser architecture, training procedure, and performance metrics.
Section~\ref{sec:results} reports the results.
We discuss their implications and limitations in
Section~\ref{sec:discussions} and summarize our conclusions in
Section~\ref{sec:conclusions}.

\section{Training data and preprocessing}
\label{sec:descript-of-simulated-data}

\subsection{Air-shower and radio-emission simulations}
\label{subsec:zhaires-sims}

We simulate the intrinsic radio emission of extensive air showers with \textsc{ZHAireS}~\cite{Alvarez-Muniz:2012, Sanchez:2025ksf}. The sample contains downward-going proton- and iron-induced showers with energies $0.4$--$4~\mathrm{EeV}$, zenith angles $37^{\circ} \le \theta \le 87^{\circ}$, and uniform azimuthal coverage. We adopt the geomagnetic field and atmospheric profile of \emph{GRANDProto300}~\cite{Guelfand:2025ncg}. The \textsc{ZHAireS} electric-field waveform defines the incident signal $v_{\rm sig}(t)$ used to construct the detector-level reference and noisy traces.

\subsection{Galactic and instrumental noise model}
\label{subsec:noise-model}

A GRAND-like detection unit passes the antenna voltage through an analog radio-frequency (RF) chain containing amplifiers, cables, filters, and digitization electronics~\cite{GRAND:2024atu}. The net transfer function modifies the pulse spectrum, defines the analysis band, and introduces a propagation delay. Diffuse Galactic emission contributes to the antenna voltage noise~\cite{GRAND:2024atu}.

For the training sample, we replace the full component-level implementation in \texttt{GRANDLib}~\cite{GRAND:2024atu} with a reduced RF-chain model. A fixed phenomenological transfer function approximates the net gain and band limitation of the low-noise amplifier (LNA), cable, and filter chain. Each trace also includes an independent timing offset of a few nanoseconds and broadband Gaussian noise normalized to the expected Galactic-noise scale (Cf. Fig.~\ref{fig:pipeline}). The model excludes the LST-dependent colored sky background, measured RFI, and time-dependent detector conditions.

For an incident signal $v_{\rm sig}(t)$, the noise-free reference and contaminated input are
\begin{align}
x_{\rm true}(t) &= \mathcal{R}\!\left\{v_{\rm sig}(t-\delta t)\right\}, \\
x_{\rm noisy}(t) &= \mathcal{R}\!\left\{v_{\rm sig}(t-\delta t)+n(t)\right\}.
\end{align}
Here, $\mathcal{R}$ denotes the reduced RF-chain response, $\delta t$ the per-trace timing offset, and $n(t)$ the additive broadband-noise realization. The first expression defines the training target, and the second defines the network input. The resulting sample tests antenna-level denoising and is not a full GRAND detector simulation.

\subsection{Dataset construction}
\label{subsec:data}

The dataset contains 410{,}673 three-polarization traces $(X,Y,Z)$, each stored with 1{,}024 samples per channel at a sampling interval of $0.5~\mathrm{ns}$. The network is trained on 512-sample windows extracted from these traces. During training, random pulse-position augmentation prevents the network from exploiting a fixed pulse location. The Ray Tune training procedure partitions the data into training, validation, and unused remainder subsets within each trial, as described in Sec.~\ref{subsec:hyperparameter-search}. The network maps $x_{\rm noisy}(t)$ to the reconstruction $x_{\rm rec}(t)$ and is trained against $x_{\rm true}(t)$.

Figure~\ref{fig:training-distribution} shows the input-SNR distribution of the evaluation sample. Most traces lie near threshold, where noise can displace the reconstructed pulse time and distort amplitude-sensitive quantities. We report timing recovery, peak-amplitude bias, and waveform fidelity separately because these quantities respond differently to denoising. At high SNR, the pulse is already resolved, so denoising has a smaller effect.

\begin{figure}[t!]
    \centering
    \includegraphics[width=1.0\linewidth]{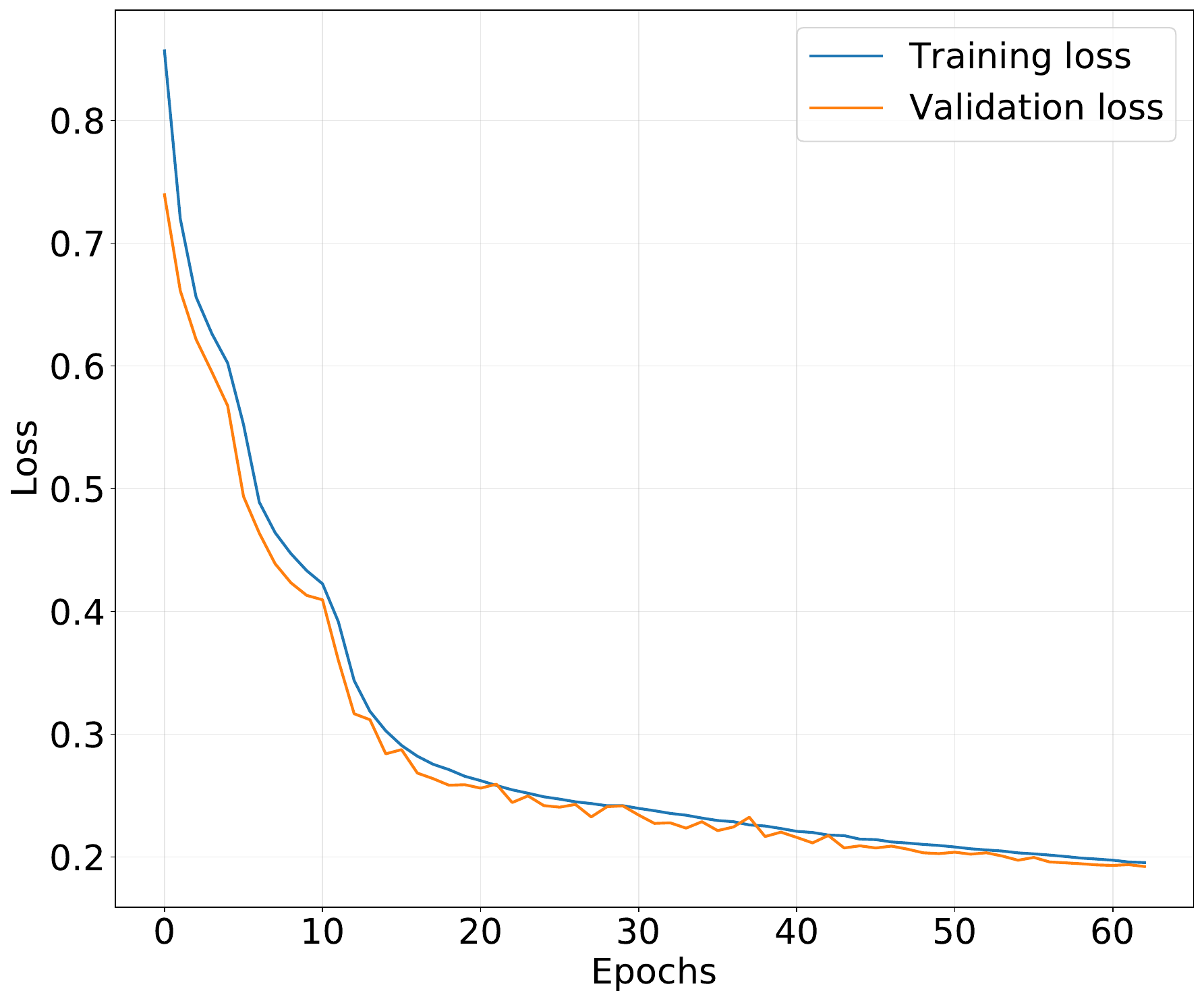}
    \caption{Training and validation values of the complete multidomain objective in Eq.~\ref{eq:multi-loss-methods} for the selected trial. Each constituent term is averaged over the batch, channels, and its time or frequency bins before the weighted sum is formed. The selected ASHA trial completed 63 of the maximum 100 epochs.}
    \label{fig:training_metrics}
\end{figure}

\begin{table*}[t!]
\caption{Selected configuration of the fiducial multidomain denoiser. The effective decoder widths account for replacement of the unused first nominal decoder entry by the fusion width.}
\label{tab:best_hyperparameters}
\begin{ruledtabular}
\begin{tabular}{lc}

Parameter & Selected value \\
\hline
Objective & Time + magnitude + direct principal-phase $L_1$ \\
Input window & $3\times512$ samples \\
Time stem channels & 64 \\
Time residual channels & $(128,256)$ \\
Magnitude/phase stem channels & 64 \\
Magnitude/phase residual channels & $(128,256)$ \\
Fusion width & 384 \\
Effective decoder widths & $(384,64,32,3)$ \\
Magnitude-loss weight & 0.011676 \\
Phase-loss weight & 0.028893 \\
Batch size & 1024 \\
Base / maximum learning rate & $1.656\times10^{-5}$ / $5.067\times10^{-4}$ \\
Cyclic step size and mode & 10{,}000; triangular2 \\
Weight decay & $2.566\times10^{-4}$ \\
Completed / maximum epochs & 63 / 100 \\
Ray Tune trials & 36 \\
\end{tabular}
\end{ruledtabular}
\end{table*}

\begin{figure*}[t!]
    \centering
    \includegraphics[width=1.0\linewidth]{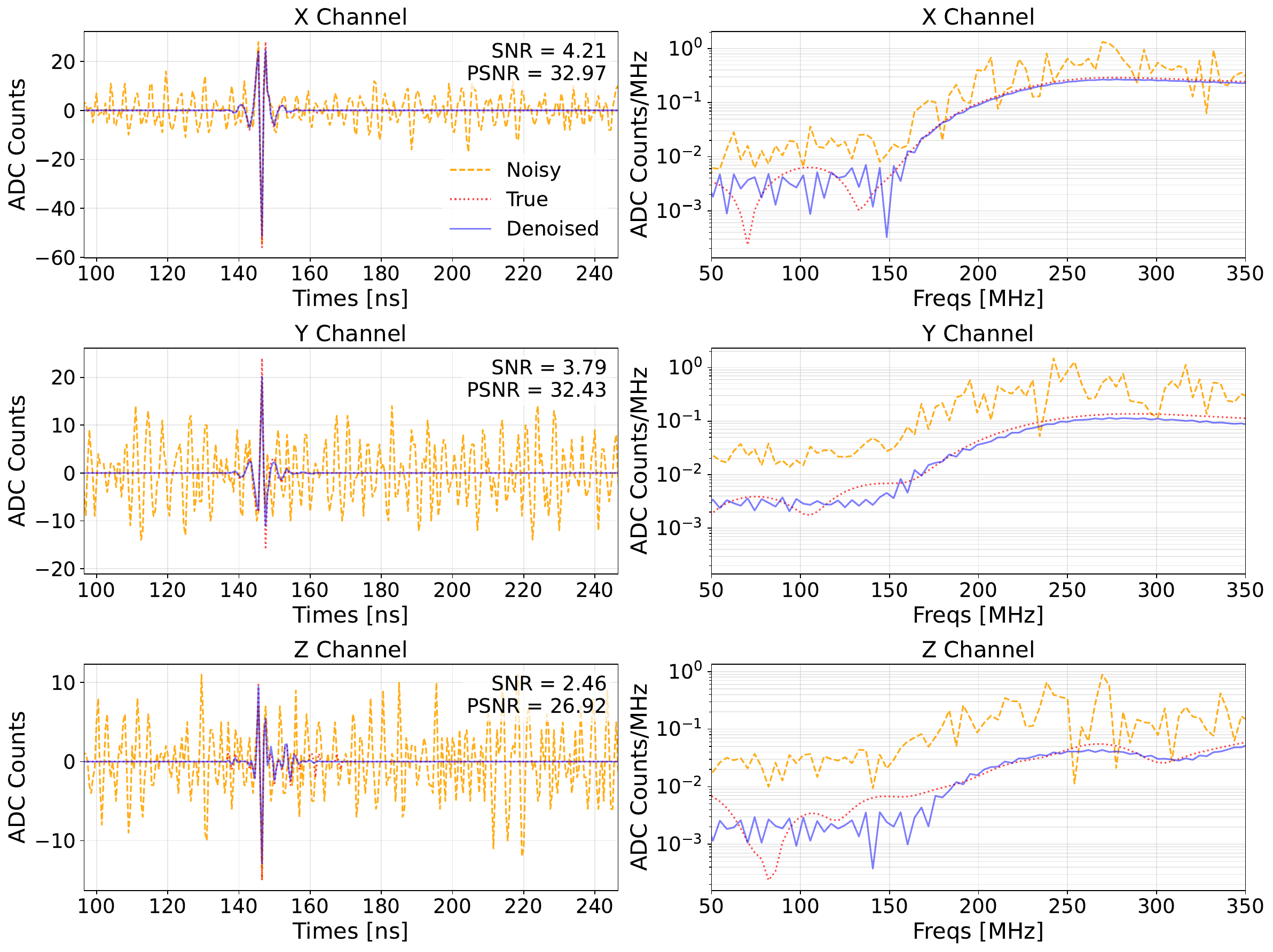}
    \caption{\textbf{Representative reconstructions in the time and frequency domains.} The rows show different antenna traces in the X, Y, and Z channels, with high, intermediate, and low input SNR, respectively. The left panels show the clean target (red dotted), noisy input (orange dashed), and denoised reconstruction (blue solid) in the time domain. The right panels show the corresponding Fourier-amplitude spectra. The time and frequency axes use the simulation sampling interval, $\Delta t=0.5~\mathrm{ns}$. The annotations give the per-channel input SNR defined in Eq.~\ref{eq:snr-methods} and the PSNR defined in Eq.~\ref{eq:psnr-methods}.}
    \label{fig:denoised-traces-time-freq}
\end{figure*}

\section{Denoiser architecture and training}
\label{sec:methods}

\subsection{Denoiser architecture}
\label{subsec:model-architecture}

Figure~\ref{fig:Model} shows the denoiser architecture. The network processes the three polarization traces jointly through time-domain and Fourier-domain pathways~\cite{Vincent:2008}. The time-domain pathway extracts local pulse structure, while the Fourier-domain pathway receives the magnitude and principal phase of the one-sided real Fourier transform of the 512-sample input window. No taper is applied before the transform.

Each encoder branch contains a stem one-dimensional convolution, two residual blocks~\citep{He:2016}, and two max-pooling operations, giving a net downsampling factor of four. The time, magnitude, and phase branches each end with 256 channels. Concatenating their outputs produces 768 channels, which the fusion module reduces to 384. A three-stage transposed-convolution decoder with effective widths $(384,64,32,3)$ reconstructs the three polarization waveforms. Table~\ref{tab:best_hyperparameters} lists the selected configuration, and Appendix~\ref{appdx:architecture} gives further implementation details. An earlier public release of the code is archived on Zenodo (DOI: \href{https://doi.org/10.5281/zenodo.18233878}{10.5281/zenodo.18233878}). The implementation and analysis products used here are available in the public \href{https://github.com/grand-mother/ML_denoising/tree/raytune_lib_final_v1}{\texttt{raytune\_lib\_final\_v1}} branch.

\begin{figure*}[t!]
    \centering
    \includegraphics[width=1.0\linewidth]{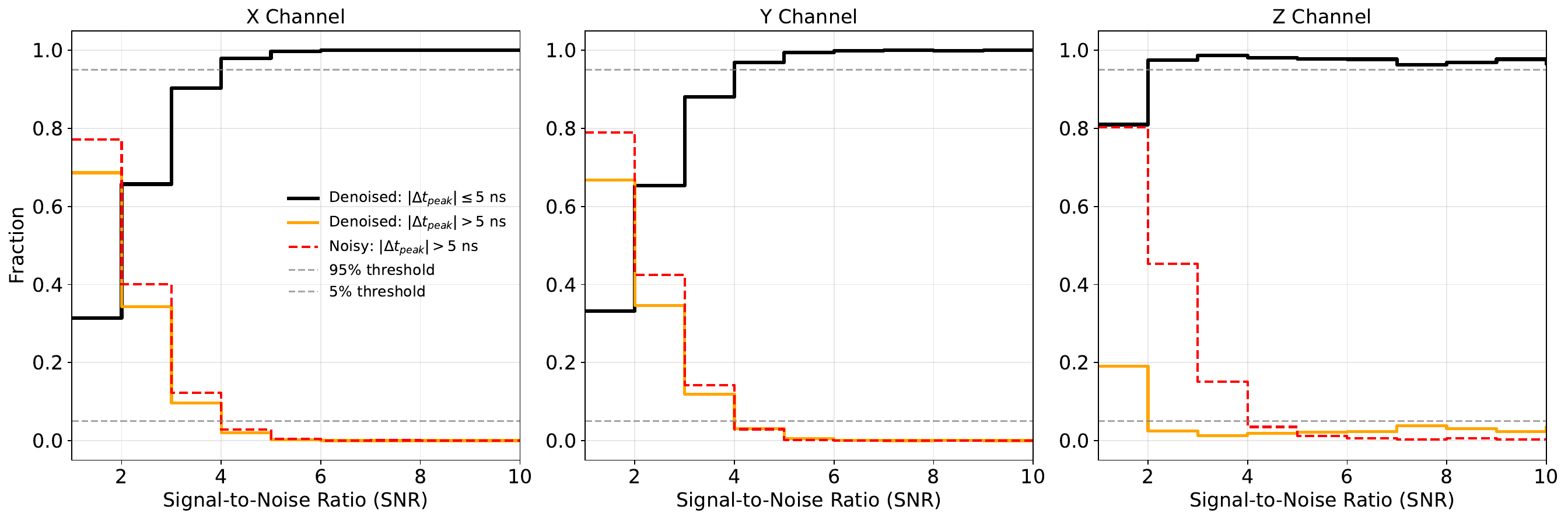}
\caption{\textbf{Timing reconstruction versus input SNR.} For each polarization channel, the black curve shows the fraction of traces passing the noisy-input preselection for which the denoised peak time agrees with the clean reference within 10 samples, or $5~\mathrm{ns}$ at the $0.5~\mathrm{ns}$ sampling interval. The orange curve shows the complementary denoised failure fraction, $|\Delta t_{\rm peak}|>5~\mathrm{ns}$, and the red dashed curve shows the corresponding failure fraction for the noisy input. Gray dashed lines mark the 95\% and 5\% levels.}

    \label{fig:peak_time_efficiency}
\end{figure*}

\subsection{Training objective}
\label{subsec:training-objective}

We train the fiducial model with a multidomain $L_1$ objective. Its time-domain term is
\begin{equation}
L_{\rm time}=\frac{1}{N}\sum_t\left|x_{\rm true}(t)-x_{\rm rec}(t)\right|.
\end{equation}
The one-sided real Fourier transforms of the reference and reconstructed traces are
\begin{equation}
\widetilde{x}_{\rm true}(f)=\mathrm{RFFT}[x_{\rm true}(t)],
\qquad
\widetilde{x}_{\rm rec}(f)=\mathrm{RFFT}[x_{\rm rec}(t)].
\end{equation}
The magnitude term is
\begin{equation}
L_{\rm mag}=\frac{1}{N_f}\sum_f
\left|\left|\widetilde{x}_{\rm true}(f)\right|-
\left|\widetilde{x}_{\rm rec}(f)\right|\right|.
\end{equation}

The fiducial loss uses the direct difference between the principal Fourier phases. With $\phi_{\rm true}=\arg\widetilde{x}_{\rm true}$ and $\phi_{\rm rec}=\arg\widetilde{x}_{\rm rec}$, the phase term is

\begin{equation}
L_{\rm phase}=\frac{1}{N_f}\sum_f
\left|\phi_{\rm rec}(f)-\phi_{\rm true}(f)\right|.
\end{equation}

The complete objective is

\begin{equation}
L_{\rm multi}=L_{\rm time}+\sigma_{\rm mag}L_{\rm mag}
+\sigma_{\rm phase}L_{\rm phase}.
\label{eq:multi-loss-methods}
\end{equation}

The weights $\sigma_{\rm mag}=0.011676$ and $\sigma_{\rm phase}=0.028893$ were selected on the validation set. The direct principal-phase difference is discontinuous at the $-\pi/\pi$ boundary. To quantify the effect of this discontinuity, we evaluated a wrapped alternative on a separate fixed sample of 41{,}067 traces by replacing the phase residual with $\operatorname{atan2}[\sin(\Delta\phi),\cos(\Delta\phi)]$ while leaving the other terms unchanged. The mean phase term changed from 2.0711 to 1.5616, whereas the complete objective changed by only $0.0043\%$. The reported analyses therefore use the checkpoint trained with the direct principal-phase convention.

\subsection{Performance metrics}
\label{subsec:metrics}

We evaluate waveform error, pulse timing, and amplitude response with separate metrics. The per-trace mean-squared error is
\begin{equation}
\mathrm{MSE}=\frac{1}{N}\sum_t\left[x_{\rm true}(t)-x_{\rm rec}(t)\right]^2.
\end{equation}
The corresponding peak signal-to-noise ratio is
\begin{equation}
\mathrm{PSNR}=10\log_{10}\!\left(\frac{A_{\rm true}^{2}}{\mathrm{MSE}}\right),
\qquad
A_{\rm true}=\max_t\left|\mathcal{H}[x_{\rm true}(t)]\right|,
\label{eq:psnr-methods}
\end{equation}
where $\mathcal{H}$ denotes the analytic-signal operation used to compute the Hilbert-envelope magnitude.

We bin the diagnostics by the simulation-level full-trace peak-to-RMS ratio
\begin{widetext}
\begin{equation}
\mathrm{SNR}_{\rm in}
=
\frac{P_{\rm true}}{\sigma_{\rm full}},
\qquad
P_{\rm true}
=
\max_t x_{\rm true}(t),
\qquad
\sigma_{\rm full}
=
\operatorname{std}_t\!\left[x_{\rm noisy}(t)\right].
\label{eq:snr-methods}
\end{equation}
\end{widetext}
Both $P_{\rm true}$ and $\sigma_{\rm full}$ are evaluated over the complete 512-sample trace. The denominator therefore includes the signal interval and is not an off-pulse estimate of the noise level. We use this convention throughout the analysis. The quantity serves only to order and bin the simulated traces and should not be compared directly with SNR values based on a separately defined noise window.

The peak-amplitude and timing diagnostics use the analytic-signal envelope over the 512-sample window. Its global maximum defines the peak amplitude and sample index for the clean, noisy, and denoised traces. We apply no additional waveform denoising to the noisy-input baseline. Appendix~\ref{appdx:wf-quality} describes the separate band-limited waveform diagnostic.

\subsection{Optimization, tuning, and data splits}
\label{subsec:hyperparameter-search}

We train the denoiser with the noisy and clean traces described in
Sec.~\ref{sec:descript-of-simulated-data}. In the original Ray Tune
search, each trial generated its own random 80/10/10 partition of the
410{,}673 traces. The training and validation subsets were used to fit
the network and assess that trial, while the remaining 10\% did not
enter the trial objective. The indices for these historical
trial-specific partitions were not retained, so the exact partition
used by the selected trial cannot now be reconstructed.

For the revised timing, peak-amplitude, and waveform-fidelity
diagnostics, we use a fixed evaluation subset generated with seed
12345. It contains 41{,}068 traces and ensures that all revised
diagnostics are evaluated on the same reproducible sample. Because the
overlap between this subset and the historical training partition of
the selected trial cannot be established, we refer to it as an
evaluation subset rather than as an independent held-out test set.

The network is implemented in \texttt{PyTorch}~\citep{Paszke:2019}
and optimized with Adam and weight decay~\citep{kingma:2017}. The
learning rate follows a triangular2 cyclical schedule~\citep{Smith:2017},
with the minimum and maximum rates and cycle length listed in
Table~\ref{tab:best_hyperparameters}.

We explore the hyperparameter space with \texttt{Ray
Tune}~\citep{Liaw:2018} and use the Asynchronous Successive Halving
Algorithm (ASHA)~\citep{Li:2020} to distribute the training budget
among the 36 trials. ASHA uses the validation loss to decide which
configurations should receive additional training, so the maximum of
100 epochs is an upper budget for a trial rather than a prescribed
training length. Under this procedure, the selected trial completed
63 epochs. Figure~\ref{fig:training_metrics} shows its training and
validation histories, Table~\ref{tab:best_hyperparameters} gives the
selected configuration, and Table~\ref{tab:hyperparam-search} lists
the ranges explored in the search.

As a separate architecture check, Appendix~\ref{appdx:time-only-comparison}
compares the dual-branch network with a time-only model. For this
comparison, both models are retrained from scratch using the same
seeded data manifest and training configuration, and are evaluated on
the same untouched test subset of 41{,}068 traces. The dual-branch
model reaches a mean test PSNR of $33.738~\mathrm{dB}$, compared with
$33.341~\mathrm{dB}$ for the time-only model; the corresponding
multidomain losses are 16.314 and 17.574. Appendix~\ref{appdx:time-only-comparison}
defines the loss and PSNR conventions used specifically for this
ablation.

\section{Results}
\label{sec:results}

A useful denoiser must do more than produce visually cleaner traces.
Near threshold, it should increase the probability of recovering a weak
pulse without increasing the false-positive rate, preserve the pulse
arrival time used in wavefront reconstruction, and retain the waveform
information relevant to amplitude-sensitive observables
(e.g., Refs.~\cite{Ferriere:2025,Decoene:2021ncf}). These requirements
probe different aspects of the reconstruction. A pulse can be recovered
at approximately the correct time even when its amplitude or detailed
waveform is imperfect. We therefore characterize three complementary
aspects of the reconstruction: detection at fixed false-positive rate,
pulse timing and peak-amplitude response, and band-limited waveform
fidelity. The latter is discussed separately in
Appendix~\ref{appdx:wf-quality}.

Figure~\ref{fig:denoised-traces-time-freq} shows representative traces
from the fiducial denoiser in the three polarization channels. At SNR
$\simeq 4.2$, the reconstructed waveform closely follows the reference
pulse while suppressing incoherent baseline fluctuations. At SNR
$\simeq 2.5$, where the pulse is less distinct in the noisy trace,
the network recovers a compact feature near the injected pulse time and
suppresses much of the broadband noise, although the reconstructed peak
amplitude is attenuated. The frequency-domain panels show the
corresponding reduction in stochastic spectral power. The ensemble tests
below quantify the same behavior. Near threshold, the largest gains
occur in pulse recovery and timing, whereas the amplitude response
requires separate evaluation.

\begin{figure*}[t!]
    \centering
    \includegraphics[width=0.95\linewidth]{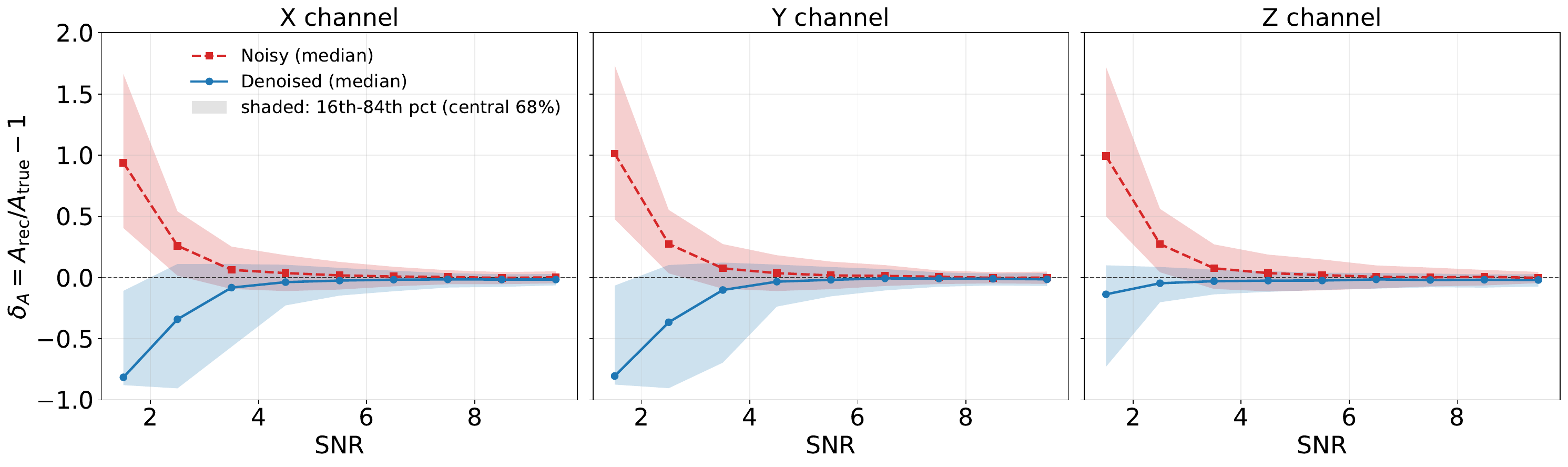}
\caption{\textbf{Peak-amplitude bias versus input SNR.}
Median signed bias, $\delta_A=A_{\rm rec}/A_{\rm true}-1$, binned in
input SNR for trace--channel pairs from the fixed evaluation subset.
The displayed unit-width bins span $1<\mathrm{SNR}_{\rm in}<10$;
bins containing fewer than 20 trace--channel pairs are omitted.
The noisy and denoised curves use identical trace--channel pairs and
SNR values, with no trigger-like selection applied. Shaded bands show
the central 68\% population interval (16th--84th percentiles).
At low SNR, the noisy peak is biased high by maximization over noise
fluctuations, whereas the denoised peak is biased low. The attenuation
decreases as the pulse becomes better resolved, although a residual
channel-dependent bias remains.}
    \label{fig:amp_bias}
\end{figure*}

\subsection{Timing recovery and amplitude response}

Timing and amplitude can respond differently to denoising.
For direction reconstruction, the relevant quantity is the pulse arrival
time at each antenna, whereas amplitude-sensitive observables depend on
the fidelity of the reconstructed electric-field scale
(e.g., Refs.~\cite{Ferriere:2025,Decoene:2021ncf}). Near threshold,
these requirements separate because the denoiser can recover the
arrival time of a weak pulse even when its reconstructed amplitude
remains imperfect.

We first examine pulse timing. To compare the noisy and denoised traces
for the same population of pulse candidates, we apply a common
trigger-like preselection based only on the noisy waveform. Let
$e_{\rm noisy}=|\mathcal{H}[x_{\rm noisy}]|$ and let $A_{\rm noisy}$
denote its maximum. We estimate the local envelope scale from the median
absolute deviation after excluding 32 samples on either side of this
maximum,
\begin{equation}
\sigma_{\rm env}
=
1.4826\,\mathrm{median}
\left\{
\left|
e_{\rm noisy}-\mathrm{median}(e_{\rm noisy})
\right|
\right\},
\end{equation}
where the medians are evaluated over the remaining samples. We retain
traces satisfying $A_{\rm noisy}\geq2\sigma_{\rm env}$. This selection
only defines a common sample for the timing comparison.

We regard the pulse time as recovered when the maximum of the denoised
Hilbert envelope lies within 10 samples of the clean reference,
corresponding to $5~\mathrm{ns}$ at the $0.5~\mathrm{ns}$ sampling
interval. Figure~\ref{fig:peak_time_efficiency} shows that denoising
reduces large timing errors at low and intermediate SNR. The improvement
decreases as the pulse emerges above the background. Once the signal
dominates the trace, the noisy waveform already identifies the peak time
reliably, and the timing efficiency approaches unity.

The amplitude response behaves differently. We do not apply the
trigger-like timing selection for this comparison. Instead, we compare
the noisy and denoised peak amplitudes for the same trace--channel pairs
in the fixed evaluation subset and bin them by the common
$\mathrm{SNR}_{\rm in}$. Figure~\ref{fig:amp_bias} shows two competing
effects near threshold. In the noisy trace, the largest envelope
excursion is often set by a positive noise fluctuation, biasing the
measured peak upward. Denoising suppresses these fluctuations, but for
the weakest pulses it also removes part of the signal and shifts the
reconstructed peak amplitude low. This attenuation decreases with
increasing SNR, although a smaller channel-dependent bias remains over
part of the displayed range.

The two diagnostics therefore separate near threshold. Denoising
improves pulse-time recovery in a regime where the raw waveform often
fails to localize the pulse, while the reconstructed amplitude retains
an SNR- and channel-dependent response. Translating the timing gain into
improved shower-direction reconstruction will depend on which antennas
are recovered, their spatial configuration, and the distribution of
timing residuals across the event.

\subsection{Detection at fixed false-positive rate}
\label{subsec:false-positives}

To estimate detection probability at a fixed false-positive rate, we construct signal-free traces from off-pulse windows. For each event, we locate the clean pulse, mask a conservative interval around it, and draw segments from the remaining contaminated trace to assemble a full-length noise-only waveform. A candidate identified in this sample is counted as a false positive~\cite{VanTrees2002}.

\begin{figure}[t!]
    \centering
    \includegraphics[width=1.0\linewidth]{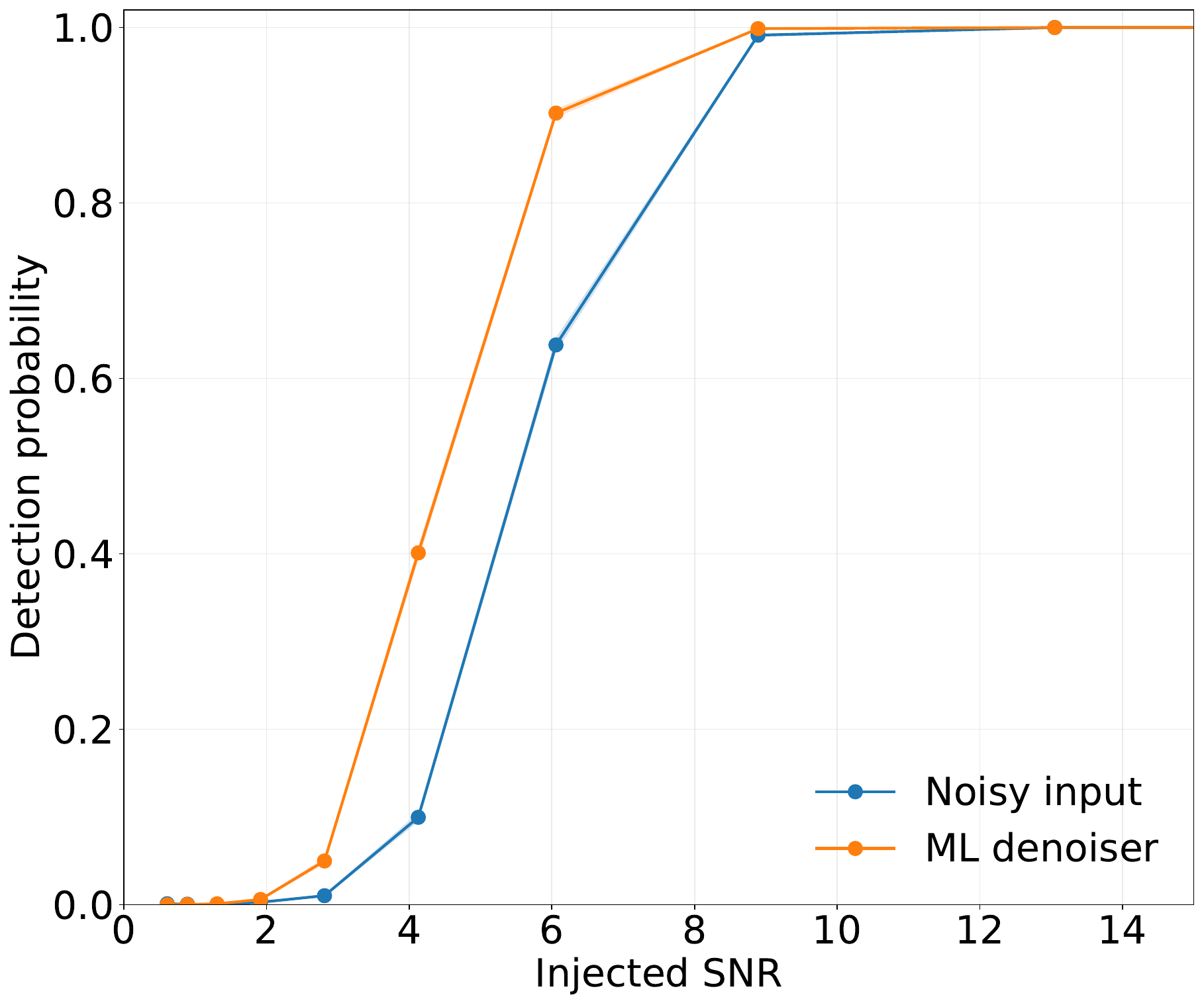}
    \caption{\textbf{Detection probability versus injected SNR at fixed
false-positive rate, $P_{\rm FP}=10^{-3}$.}
The blue and orange curves show candidate selection on the noisy and
ML-denoised traces, respectively. For each method, the threshold is
calibrated using its own constructed noise-only sample to enforce the
same false-positive rate. The comparison therefore measures the gain
in detection probability under the simulated broadband-noise background
considered here.}
    \label{fig:falsepositives}
\end{figure}

We calibrate separate candidate thresholds for the noisy input and denoised output using their respective noise-only samples. Each threshold is chosen to give one false positive per 1{,}000 traces, $P_{\rm FP}=10^{-3}$, and is then applied to signal-containing traces. Figure~\ref{fig:falsepositives} shows the resulting detection probability as a function of injected SNR. At the same false-positive rate, the denoiser recovers more signal-containing traces, with the largest difference near threshold.

\section{Discussion}
\label{sec:discussions}

The clearest effect of denoising is the recovery of pulse timing near
threshold. Planar wavefront reconstruction depends on the relative
arrival times of the radio pulse across the array and on the antenna
positions~\cite{Ferriere:2025}. Recovering a reliable timestamp from a
trace in which the pulse is otherwise difficult to localize can
therefore contribute directly to the wavefront fit. The resulting
improvement in shower-direction reconstruction will depend on the
spatial configuration of the recovered antennas and the distribution
of their timing residuals, rather than simply on the number of recovered
antennas.

The amplitude response exhibits a different behavior near threshold.
In the noisy trace, maximizing the envelope tends to select positive
noise fluctuations and can bias the measured peak upward. The denoiser
removes much of this stochastic contribution, but for the weakest pulses
it also suppresses part of the signal, producing the negative amplitude
bias seen in Fig.~\ref{fig:amp_bias}. This attenuation decreases as the
pulse becomes better resolved, although a smaller channel-dependent
response remains at higher SNR. Timing and amplitude therefore separate
in the noise-dominated regime. In particular, the pulse arrival time can be recovered
even when its amplitude remains biased.

This distinction is particularly relevant for energy-sensitive radio
observables. The measured electric field is a superposition of signal
and noise, and random relative phases can produce constructive or
destructive interference at low SNR~\cite{martinelli:2025}. Amplitude
and fluence reconstruction therefore require explicit treatment of the
noise contribution even without denoising. Here, the network introduces
an additional amplitude response that we measure as a function of SNR
and polarization. Characterizing this response is a first step toward
calibrating amplitude-sensitive observables.

The fixed-false-positive comparison gives a complementary view of the
near-threshold gain. At $P_{\rm FP}=10^{-3}$, the denoised traces have a
higher detection probability than the noisy traces, with the largest
difference where the pulse is only weakly resolved. The improvement in
timing is therefore accompanied by better separation of signal-containing
and noise-only traces under the simulated broadband background considered
here.

The matched time-only comparison in Appendix~\ref{appdx:time-only-comparison}
provides a separate check of the role of the Fourier pathways. Under the
same training configuration, the dual-branch model improves the mean test
PSNR by $0.40~\mathrm{dB}$ and gives a lower multidomain loss than the
time-only model. The effect is modest, but both metrics favor the
dual-branch architecture. We therefore view the frequency-domain pathways
as a useful component of the denoiser rather than the main source of its
performance.

\section{Conclusions}
\label{sec:conclusions}

We developed a convolutional denoiser for GRAND-like radio traces that combines time-domain information with
Fourier-domain magnitude and phase. In the simulations considered
here, denoising increases the detection probability at fixed
false-positive rate and improves pulse-time recovery near threshold.
In particular, the network recovers timing information from traces in
which the pulse is otherwise difficult to localize. Because relative
pulse times are the basic input to radio-wavefront reconstruction, this
recovery increases the timing information available for air-shower
reconstruction.

The amplitude response behaves differently. For weak pulses, denoising
suppresses stochastic fluctuations but also attenuates part of the
signal, producing a negative peak-amplitude bias. This attenuation
decreases as the pulse becomes better resolved, while a smaller
channel-dependent response persists at higher SNR. Together with the
band-limited waveform test, these results show that pulse timing and
overall waveform recovery can improve before the peak amplitude reaches
comparable fidelity. The measured SNR- and polarization-dependent
amplitude response provides a basis for calibrating amplitude-sensitive
observables.

For the present denoiser, candidate recovery and timing show the largest
near-threshold improvements, while the amplitude response must be
accounted for in energy-sensitive reconstruction. The next step is to
test these gains with more complete GRAND detector simulations and
measured background conditions, and to propagate the recovered pulse
information through shower-direction and energy reconstruction.

\section*{Data and code availability}
The code repository is available on GitHub (\href{https://github.com/grand-mother/ML_denoising}{grand-mother/ML\_denoising}). The analysis scripts, evaluation manifests, figure-provenance records, and figures used in this study are available in the public \href{https://github.com/grand-mother/ML_denoising/tree/raytune_lib_final_v1}{\texttt{raytune\_lib\_final\_v1}} branch. The earlier public release is archived on Zenodo (DOI: \href{https://doi.org/10.5281/zenodo.18233878}{10.5281/zenodo.18233878}). The clean waveforms are generated with \textsc{ZHAireS}. Section~\ref{sec:descript-of-simulated-data} describes the detector-response and noise model. Additional data products are available from the authors upon reasonable request.

\begin{acknowledgments}
We thank Sara El Bouch, Claire Gu\'epin, and Simon Prunet for useful discussions. We also thank the GRAND Collaboration for providing the \emph{GRANDProto300} site parameters and related technical information used in this study. OM and ZL acknowledge support from the U.S. National Science Foundation under Grant No. 2418730.
\end{acknowledgments}

\bibliography{zreferences}

\appendix

\section{Band-limited waveform-fidelity diagnostic}
\label{appdx:wf-quality}

We test whether identical band conditioning leaves the denoised output closer to the injected waveform than the noisy input. The fiducial model is evaluated on a fixed seed-12345 evaluation subset of 41{,}068 traces. Because the historical training partition of the selected Ray Tune trial was not archived, we do not assume that this evaluation subset is disjoint from that trial's training data. This is a waveform-reconstruction diagnostic, not a calibration of amplitude, fluence, or primary energy.

We use the sampling interval of the simulated traces, $\Delta t=0.5~\mathrm{ns}$, and apply a fourth-order Butterworth band-pass over $\mathcal{B}=[50,200]~\mathrm{MHz}$ to the reference, noisy, and reconstructed traces~\cite{GRAND:2024atu}. The nominal fidelity interval is centered on the clean Hilbert-envelope maximum and extends by $150~\mathrm{ns}$ on either side. At this sampling interval, the nominal interval spans 601 samples, which is wider than the 512-sample evaluation trace. The interval is therefore clipped at the trace boundaries, and the fidelity metrics are evaluated over the full 512-sample trace. We denote this effective interval by $\mathcal{W}$ below. For each polarization channel, the normalized mean-squared error is
\begin{equation}
\mathrm{NMSE}_{\mathcal B}=
\frac{\sum_{t\in\mathcal{W}}\left[x^{\mathcal B}_{\rm rec}(t)-x^{\mathcal B}_{\rm true}(t)\right]^2}
{\sum_{t\in\mathcal{W}}\left[x^{\mathcal B}_{\rm true}(t)\right]^2}.
\end{equation}
The noisy-input baseline is obtained by replacing $x^{\mathcal B}_{\rm rec}$ with the band-passed noisy trace over the same interval $\mathcal{W}$.

\begin{figure*}[t!]
    \centering
    \includegraphics[width=1.0\linewidth]{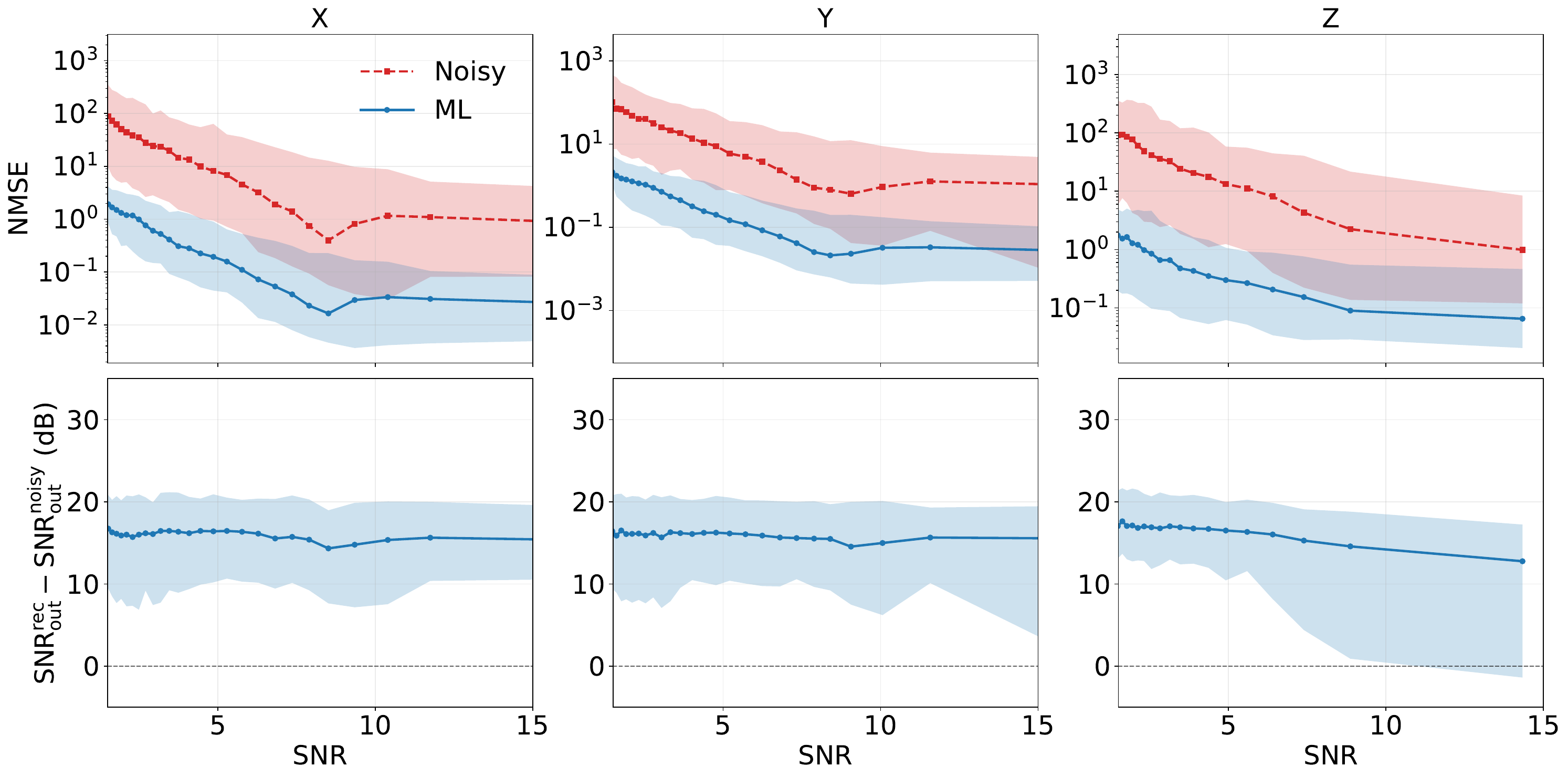}
    \caption{\textbf{Band-limited waveform fidelity.} The top row shows the rolling median $\mathrm{NMSE}_{\mathcal B}$ for the ML reconstruction and the band-passed noisy-input baseline after applying the fourth-order $50$--$200~\mathrm{MHz}$ Butterworth band-pass. The bottom row shows $\Delta\mathrm{SNR}_{\rm out}=\mathrm{SNR}_{\rm out}(x^{\mathcal B}_{\rm rec})-\mathrm{SNR}_{\rm out}(x^{\mathcal B}_{\rm noisy})$. Both quantities are evaluated over the full 512-sample, band-limited trace and binned using the full-trace input SNR in Eq.~\ref{eq:snr-methods}. The nominal $\pm150~\mathrm{ns}$ fidelity interval is wider than the trace at $\Delta t=0.5~\mathrm{ns}$ and is therefore clipped to the available samples. Shaded bands contain the central 90\% population interval in the rolling windows. A truth-dependent clean-power gate removes the lowest 10\% of positive clean powers in each channel. Lower NMSE or higher output SNR indicates closer band-limited agreement with the reference waveform but does not establish an unbiased amplitude or fluence estimator.}
    \label{fig:nmse_snrgain_vs_snr}
\end{figure*}

For a candidate waveform $y\in\{x^{\mathcal B}_{\rm noisy},x^{\mathcal B}_{\rm rec}\}$, we define
\begin{equation}
\mathrm{SNR}_{\rm out}(y)=10\log_{10}\!\left[
\frac{\sum_{t\in\mathcal W}\left(x^{\mathcal B}_{\rm true}(t)\right)^2}
{\sum_{t\in\mathcal W}\left(y(t)-x^{\mathcal B}_{\rm true}(t)\right)^2}
\right].
\end{equation}
The lower panels of Fig.~\ref{fig:nmse_snrgain_vs_snr} show
$\Delta\mathrm{SNR}_{\rm out}=\mathrm{SNR}_{\rm out}(x^{\mathcal B}_{\rm rec})-\mathrm{SNR}_{\rm out}(x^{\mathcal B}_{\rm noisy})$. To avoid ill-conditioned ratios in channels with negligible injected power, we exclude trace--channel pairs below the 10th percentile of positive clean power in $\mathcal W$, separately for each channel. This truth-dependent gate is applied identically to the noisy and denoised curves and is used only for this simulation diagnostic.

Figure~\ref{fig:nmse_snrgain_vs_snr} shows lower band-limited NMSE and a positive output-SNR difference for the ML reconstruction over much of the displayed range. The largest change occurs in the noise-dominated regime. As Fig.~\ref{fig:amp_bias} shows, however, a lower integrated waveform error can coexist with a biased peak amplitude.

\section{Autoencoder architecture}
\label{appdx:architecture}

The network maps a noisy window $x_{\rm noisy}\in\mathbb{R}^{3\times512}$ to a reconstruction $x_{\rm rec}$ of the same shape. The time-domain pathway begins with a one-dimensional convolution with 64 output channels, followed by residual stages with 128 and 256 channels. The magnitude and phase pathways use the same channel progression. Two max-pooling operations in each branch give a downsampling factor of four.

The final 256-channel outputs of the time, magnitude, and phase branches are concatenated to form 768 channels. The fusion module reduces this representation to 384 channels. The effective decoder widths are $(384,64,32,3)$. The fusion width overwrites the first nominal decoder entry in the Ray Tune configuration, so this entry is not an independent model parameter. Transposed convolutions restore the 512-sample length and return the three reconstructed polarization traces.

\begin{table*}[t!]
\centering
\caption{Hyperparameter search ranges for the dual-branch denoiser.}
\label{tab:hyperparam-search}
\begin{ruledtabular}
\begin{tabular}{ll}
\hline
\textbf{Parameter} & \textbf{Search space} \\
\hline
Time-branch stem channels & $\{64,128\}$ \\
Time-branch residual channels & $\{(64,128),(128,256)\}$ \\
Frequency-branch stem channels & $\{64,128\}$ \\
Frequency-branch residual channels & $\{(64,128),(128,256)\}$ \\
Nominal decoder channels$^{a}$ & $\{(128,64,32,3),(256,128,64,3),$ \\
& $\quad(64,32,16,3),(32,16,8,3)\}$ \\
Base learning rate & log-uniform in $[10^{-6},10^{-4}]$ \\
Maximum learning rate & log-uniform in $[10^{-5},10^{-3}]$ \\
Cyclic step size & $\{10{,}000,20{,}000,30{,}000,40{,}000,50{,}000\}$ \\
Learning-rate mode & $\{\mathrm{triangular},\mathrm{triangular2}\}$ \\
Batch size & $\{512,1024,2048,4096\}$ \\
Weight decay & log-uniform in $[10^{-5},5\times10^{-3}]$ \\
Magnitude-loss weight & log-uniform in $[10^{-2},1]$ \\
Phase-loss weight & log-uniform in $[10^{-2},1]$ \\
\hline
\end{tabular}
\end{ruledtabular}
\end{table*}

$^{a}$The fiducial model replaces the first nominal decoder entry with the fusion width. Only the remaining decoder entries affect the instantiated model.
\section{Training and validation}
\label{appdx:training}

We implement the model in \texttt{PyTorch}~\citep{Paszke:2019} and optimize it with Adam and weight decay~\citep{kingma:2017}. The learning rate follows a triangular2 cyclic schedule~\citep{Smith:2017}. We select hyperparameters with \texttt{Ray Tune}~\citep{Liaw:2018} and ASHA~\citep{Li:2020} over 36 trials with a maximum budget of 100 epochs. The selected trial completed 63 epochs. Each historical Ray Tune trial generated its own random training and validation partitions; the exact indices used by the selected trial were not archived, as discussed in Sec.~\ref{subsec:hyperparameter-search}.

\section{Time-only comparison of the denoiser architecture}
\label{appdx:time-only-comparison}

We compare the dual-branch model with a time-only model under the same training configuration. The two models use the same 512-sample inputs, seeded 80/10/10 data manifest, preprocessing, pulse-position augmentation, channel widths, fusion implementation, optimizer, cyclic learning-rate schedule, batch size, and 100-epoch training budget. They differ only in whether the Fourier-magnitude and phase pathways are enabled. No additional hyperparameter search is performed.

For this ablation, we use the selected channel widths and optimizer hyperparameters and retrain both models from scratch. Both models are trained with the wrapped angular phase residual
\begin{equation}
\Delta\phi_{\rm wrap}
=
\operatorname{atan2}\!\left[\sin(\phi_{\rm rec}-\phi_{\rm true}),
\cos(\phi_{\rm rec}-\phi_{\rm true})\right],
\end{equation}
rather than the direct principal-phase residual used by the fiducial model. The ablation loss is therefore
\begin{equation}
L_{\rm ab}=L_{\rm time}+\sigma_{\rm mag}L_{\rm mag}
+\sigma_{\rm phase}L_{\rm phase}^{\rm wrap}.
\end{equation}
Because the same objective is used for both ablation models, the phase convention does not favor either architecture.

We also record the ablation PSNR using the trace-level convention for this comparison. For trace $i$, containing all three polarization channels, we define
\begin{equation}
\mathrm{PSNR}_{{\rm ab},i}
=
20\log_{10}\!\left[
\frac{\max_{c,t}|x_{{\rm true},i,c}(t)|}
{\sqrt{\mathrm{MSE}_i}}
\right],
\end{equation}
where $\mathrm{MSE}_i$ is averaged over all channels and samples. We then average $\mathrm{PSNR}_{{\rm ab},i}$ over the test traces. This convention is specific to the ablation and should not be confused with the per-channel Hilbert-envelope PSNR in Eq.~\ref{eq:psnr-methods}.

After selecting the best checkpoint from each run on the validation subset, we evaluate both models once on the same untouched test subset of 41{,}068 traces using fixed 512-sample windows and no augmentation. The dual-branch model gives $L_{\rm ab}=16.314$ and $\langle\mathrm{PSNR}_{\rm ab}\rangle=33.738~\mathrm{dB}$, compared with $L_{\rm ab}=17.574$ and $33.341~\mathrm{dB}$ for the time-only model. Thus, the time-only loss is 7.7\% larger, while the Fourier pathways increase the mean test PSNR by $0.40~\mathrm{dB}$. This paired comparison indicates a modest improvement from the Fourier pathways. With one training realization per model, it does not estimate run-to-run statistical uncertainty.

\end{document}